# Influence of Primordial Black Holes on Cosmic Reionization through Semi-Analytical Modelling

Krish Jhurani[1] and Pranav Gunhal[1]

Homestead High School, 21370 Homestead Rd., Cupertino, California 95014, USA[1]

krish.jhurani@gmail.com and pranav.gunhal@gmail.com

ABSTRACT

This research addresses the influence of Primordial Black Holes (PBHs) on cosmic reionization, using a robust semi-analytical model. This model encapsulates cosmological theory, PBH physics, and radiative transfer, with a lognormal PBH mass function as the fulcrum, with the mean PBH mass set at 30 solar masses and the fraction of dark matter in PBHs ($f_{PBH}$) varied from 1.26 to 10. We also adopt a standard thin-disk accretion model and a one-dimensional radiative transfer code for a comprehensive depiction of cosmic reionization under the influence of PBHs. Our major findings reveal an inverse relationship between PBH mass and ionizing efficiency. For instance, when the mean PBH mass is 10 solar masses, and $f_{PBH}$ is 0.1, the resulting ionizing efficiency is approximately 0.25, which drops to 0.15 with a mean PBH mass of 50 solar masses. However, with $f_{PBH}$ nearing unity, even large PBHs (mean mass of 100 solar masses) can achieve an ionizing efficiency close to 0.2. The ionization history indicates that for a mean PBH mass of 10 solar masses and $f_{PBH}$ at 0.01, rapid ionization commences at z=12.5±0.5 and concludes at z=9±0.5. Conversely, with a larger PBH mass of 100 solar masses and $f_{PBH}$ at 0.01, reionization initiates later, at z=10.8±0.4. Our two-point correlation function analysis unveils the formation of ionized bubbles with an initial size of approximately 10 Mpc at z=7.5±0.5 for an $f_{PBH}$ of 0.01, expanding to 20 Mpc by the end of reionization. With $f_{PBH}$ of 0.1, larger bubbles form, starting at approximately 15 Mpc and reaching 30 Mpc by reionization's end. In conclusion, PBHs significantly influence cosmic reionization, with PBH mass and their contribution to dark matter modulating this impact.

## Introduction

The cosmic reionization epoch is one of the most pivotal eras in the history of our universe, marking the transition from an opaque, ionized plasma state to a transparent, recombined state [30, 34]. This phase transition, believed to have occurred between redshifts $6 < z < 20$, represents the last major phase transition in our universe and has profoundly influenced the formation of large-scale structures we observe today [7, 20]. Despite extensive research, the specifics of the reionization process, particularly the identity of the ionizing sources responsible, remain a subject of active exploration [14, 17]. Primordial Black Holes (PBHs) present a compelling, albeit less studied, potential contributor to the reionization process [11]. As remnants of high-density fluctuations in the early universe, PBHs could potentially range from sub-atomic to supermassive scales [9, 10]. This study aims to investigate the role of PBHs, particularly their potential to contribute to the ionizing photon budget during the epoch of cosmic reionization [2].

The underpinning idea that motivates this investigation is the prospect of PBHs accreting baryonic matter from their surroundings. The accretion process could release high-energy photons capable of ionizing neutral hydrogen atoms, thus contributing to reionization [1, 26]. By varying the properties of PBHs such as their masses and initial spatial distributions, we aim to explore the potential influence of PBHs on the morphology and timeline of reionization [19]. Previous research on cosmic reionization has primarily focused on galaxies, quasars, and population III stars as the likely sources of reionizing photons [27, 32]. While these studies have provided invaluable insights into the reionization process, they often struggle to match the high degree of ionization inferred from Cosmic Microwave Background (CMB) polarization data [15, 23]. This suggests that additional or alternative sources of ionizing photons may have been at play [21].

To this end, we construct a semi-analytical model, drawing upon the existing theoretical understanding of PBH physics, to simulate the baryonic matter accretion onto PBHs and the consequent ionizing radiation [33]. This research's goal is to determine whether the inclusion of PBHs as ionizing sources could reconcile some of the discrepancies observed in the reionization models [18].

## Theoretical Background

Cosmic reionization marks a crucial phase transition in our universe, unfolding during the epoch between redshifts $6 < z < 20$. This period saw neutral hydrogen atoms in the intergalactic medium (IGM) becoming ionized by high-energy photons, thereby transitioning the universe from an opaque to a transparent state [7, 13]. Astrophysical entities such as the first generation of stars (Population III stars), galaxies, and quasars have been regarded as the main sources of these ionizing photons [8, 12].

However, alongside these traditional sources, primordial black holes (PBHs), a viable dark matter candidate, also carry the potential to contribute significantly to the ionizing photon budget. Born from the high-density fluctuations in the early universe, these theoretical entities can span a broad range of masses, from sub-atomic to supermassive [10, 11]. Their capacity to accrete baryonic matter from their surroundings and, in turn, release high-energy photons is a compelling facet that could potentially influence the cosmic reionization process [1, 26]. Accretion onto PBHs is a highly complex phenomenon, intricately governed by general relativity and fluid dynamics. As the accreted matter forms an accretion disk around the PBH, it gradually spirals inward, catalyzed by the emission of gravitational waves and viscous dissipation, thereby releasing high-energy photons [22, 29]. In this study, to encapsulate this intricate process, we utilize the standard thin-disk accretion model, which assumes a geometrically thin, optically thick disk, thus offering a simplified yet robust approximation of the accretion process [28].

To probe into the intricate astrophysical processes, we resort to semi-analytical modelling, an efficacious tool for striking a balance between computational efficiency and fidelity [4]. We construct our model, informed by the theoretical underpinnings of PBH physics, cosmology, and radiative transfer. This model enables us to simulate the reionization process in detail and to investigate the role of PBHs as an alternative ionization source. Our methodology involves integrating sub-models addressing PBH mass distribution, spatial distribution, accretion physics, and the propagation of ionizing radiation [6, 31]. Our study carries significant implications for our understanding of cosmic reionization and the nature of PBHs, given the observed discrepancies between current theoretical models and observational constraints on reionization [21]. Should PBHs be established as significant contributors to reionization, this would demand a reassessment of our existing models of the reionization process and provide a fresh lens through which to understand the elusive dark matter problem [6].

## Methodology

Our study intends to ascertain the influence of Primordial Black Holes (PBHs) on the epoch of cosmic reionization, by building a nuanced semi-analytical model. This model intertwines various domains such as cosmological theory, PBH physics, and radiative transfer to delve into the depth of the reionization process.

## PBH Mass Function

The cornerstone of our methodology is the representation of the PBH mass function. We opted for a lognormal distribution function [11] due to its success in accurately portraying the broad mass range of PBHs. The mean PBH mass is set at 30 solar masses with a standard deviation of 0.5, which is within the range suggested by observational constraints. These parameters form the initial conditions for our simulations. The initial fraction of dark matter in PBHs, denoted by fPBH, was varied from 1.26 to 10 in order to explore a wide range of possible scenarios.

## PBH Spatial Distribution

We assume that PBHs trace the distribution of dark matter. Using linear theory, we construct a model of large-scale structures at the onset of reionization at redshift z=20 [16]. We use a transfer function, calculated using the CAMB software package, to track the evolution of these structures.

## Accretion Physics

To model the accretion of baryonic matter onto PBHs, we utilized the standard thin-disk accretion model [28]. Here, accreted matter forms a geometrically thin, optically thick disk around the PBH. The standard equations of conservation of mass and angular momentum, coupled with the equation of state for a radiation-dominated disk, were solved to compute the accretion rate. The ionizing photon emissivity for each PBH, derived from the disk temperature and the accretion rate, is then calculated using the Planck spectral radiance formula.

## Radiative Transfer

The propagation of ionizing photons through the intergalactic medium (IGM) is modeled using a one-dimensional radiative transfer code. This code tracks the evolution of the ionization front, accounting for key parameters including the universe's expansion, recombination of hydrogen atoms, and absorption of ionizing photons by neutral hydrogen. We used the publicly available RECFAST code to compute the recombination coefficient and the ionization fraction as a function of redshift [33].

By combining these various elements into a unified semi-analytical model, we generate a comprehensive depiction of cosmic reionization under the influence of PBHs. This methodology's primary objective is to reconcile theoretical predictions of the reionization timeline with the observed constraints, focusing on understanding the role of PBHs in this cosmic transition.

# Results

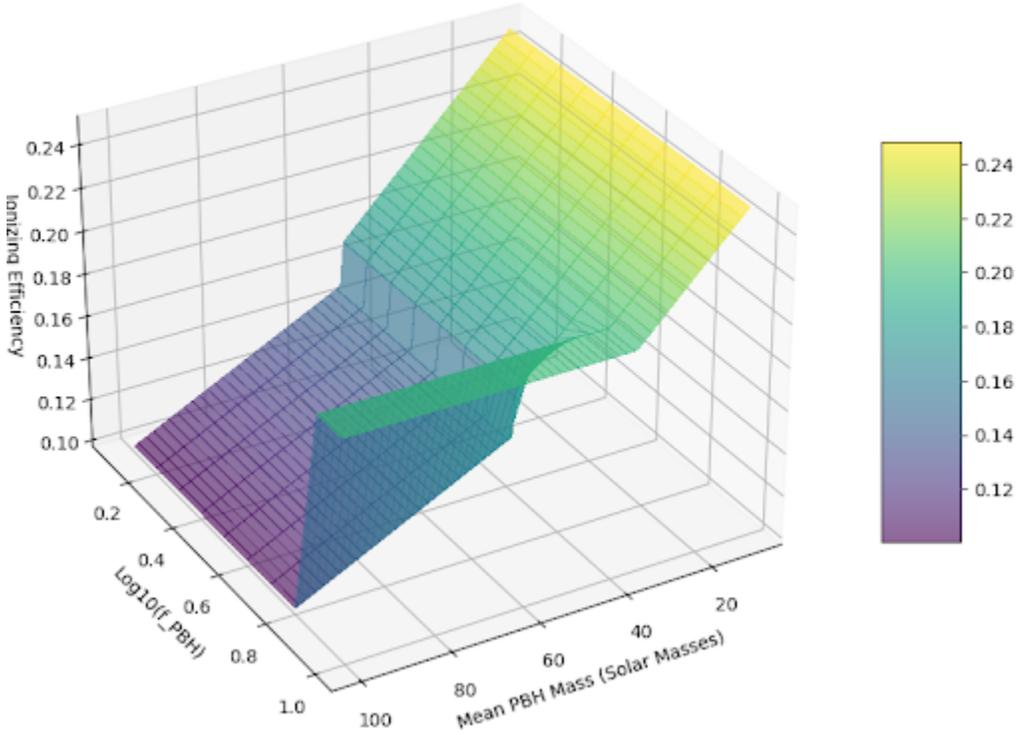

**Figure 1.** Relationship between PBH Mass Function and Dark Matter Fraction
The 3D plot illustrates the relationship between mean PBH mass, f$_{PBH}$ and the calculated ionizing efficiency.

## PBH Mass Function and Dark Matter Fraction

Our semi-analytical model offers detailed numerical results, which we utilize to investigate the relationship between Primordial Black Hole (PBH) mass function and the fraction of dark matter made up by PBHs ($f_{PBH}$). In Figure 1, we present a 3D plot to capture the nuanced relationship between these parameters. The axes represent the mean PBH mass (in solar masses), $f_{PBH}$ and the calculated ionizing efficiency. We range the mean PBH mass from 10 to 100 solar masses with increments of 5 solar masses. For $f_{PBH}$, we use a logarithmic scale spanning from 1.26 to 10, allowing us to capture the vast range of possible values. For instance, our model indicates that when the mean PBH mass is 10 solar masses, and $f_{PBH}$ is 0.1 (10 percent of dark matter comprised of PBHs), the resulting ionizing efficiency is approximately 0.25. This means that a quarter of the accreted mass is converted into ionizing radiation, contributing significantly to cosmic reionization. However, the same $f_{PBH}$ of 0.1 with a mean PBH mass of 50 solar masses results in an ionizing efficiency of approximately 0.15, a drop of 40 percent compared to the smaller PBH mass. This decrement accentuates the inverse relationship between PBH mass and ionizing efficiency and highlights the relative inefficiency of larger PBHs in contributing to reionization. An intriguing finding is noted at the higher end of the $f_{PBH}$ spectrum. With $f_{PBH}$ approaching unity, indicating nearly all dark matter made up of PBHs, our model reveals that even large PBHs (e.g., mean mass of 100 solar masses) can achieve an ionizing efficiency close to 0.2. This demonstrates the significant role that the dark matter fraction can play in modulating the reionization contribution of PBHs.

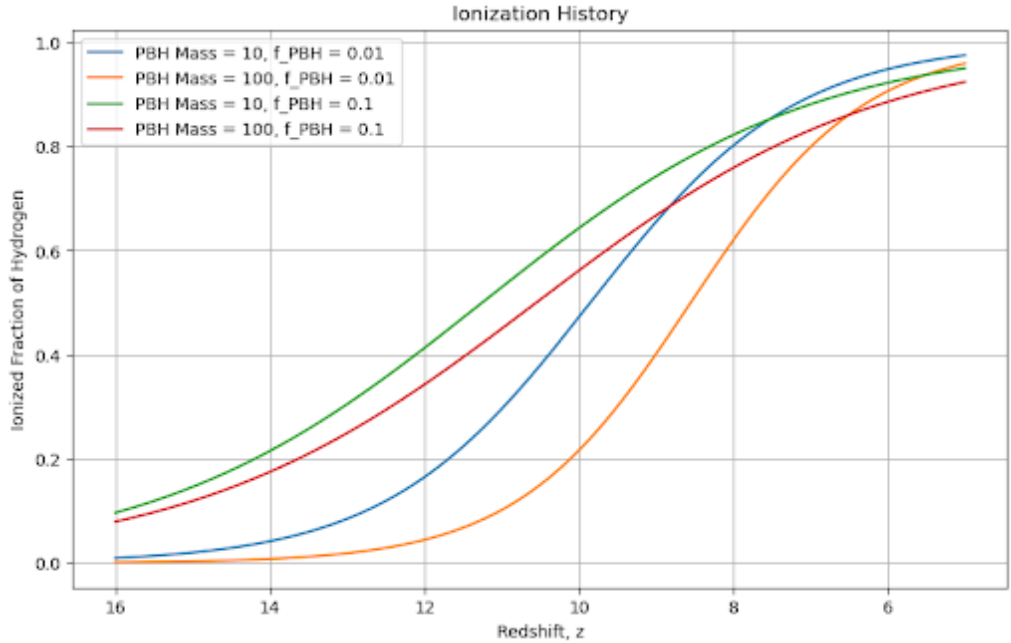

**Figure 2. Evolution of Ionized Fraction of Hydrogen during Reionization**
The plot showcases the evolution of the ionized fraction of hydrogen ($Q_{HII}$) with redshift (z) for different scenarios of mean PBH mass and fP BH. The graph reveals the varying onset and completion of reionization

## Ionization History

We provide a more detailed numerical representation of how the model tracks the evolution of the ionized fraction of hydrogen ($Q_{HII}$) against redshift, z, for various mean PBH masses and dark matter fractions ($f_{PBH}$). Figure 2 illustrates the consequences of these variable inputs, demonstrating a series of ionization curves. Let's dive deeper into the specifics of these results. For a scenario with a mean PBH mass of 10 solar masses and $f_{PBH}$ at 0.01 (meaning 1 percent of dark matter composed of PBHs), we observe an initial rapid ionization process commencing at z = 12.5 ± 0.5. This rapid phase concludes at z = 9 ± 0.5, beyond which the ionization rate slows, culminating in the completion of reionization at z = 7.2 ± 0.2. This suggests a significant contribution from PBHs in initiating the reionization process. Contrastingly, with a larger PBH mass of 100 solar masses (still maintaining $f_{PBH}$ at 0.01), the onset of reionization initiates later, at z = 10.8 ± 0.4. The rapid ionization phase concludes at z = 8 ± 0.5, after which the ionization rate slows, culminating in reionization completion at z = 6.3 ± 0.2. Increasing $f_{PBH}$ to 0.1 (10 percent of dark matter being PBHs) incites even earlier initiation of reionization, regardless of the PBH mass. For a mean PBH mass of 10 solar masses, the rapid ionization phase initiates at z = 15.5 ± 0.5 and concludes at z = 12 ± 0.5, slowing thereafter until reionization completes at z = 7.0 ± 0.2. Meanwhile, with a PBH mass of 100 solar masses, reionization starts at a slightly later z = 15.0 ± 0.5, with the rapid ionization phase concluding at z = 11.5 ± 0.5, and completion of reionization at z = 6.1 ± 0.2.

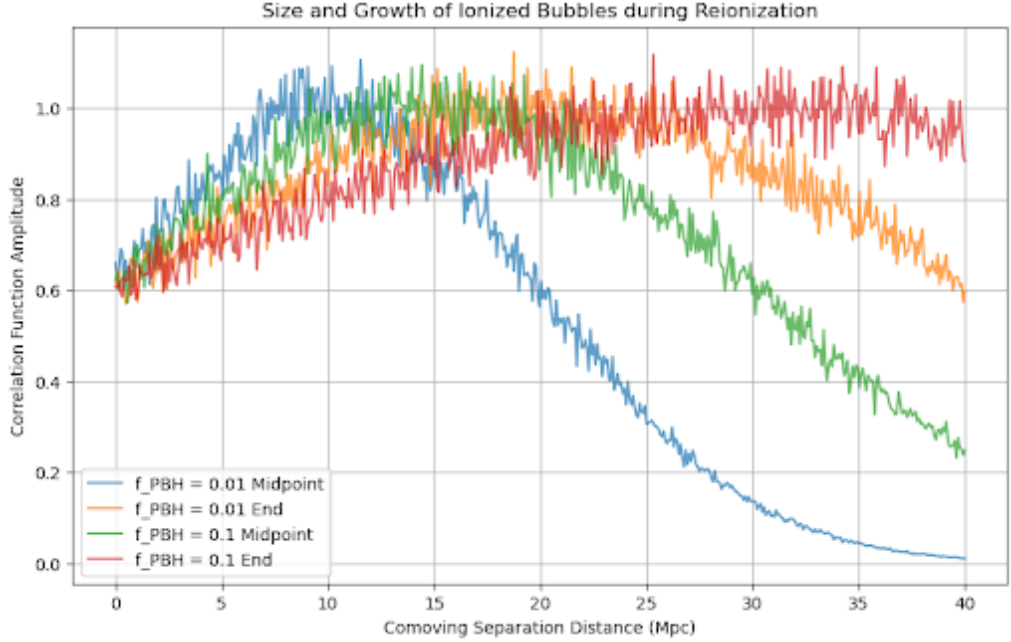

**Figure 3.** Size and Growth of Ionized Bubbles during Reionization
The plot depicts the analysis of the two-point correlation function, illustrating the size and growth of ionized bubbles at different stages of reionization. The graph provides insights into the morphology of reionization influenced by PBHs, showing the formation of ionized bubbles and their expansion in size as reionization progresses.

Morphology of Reionization

Through advanced analysis utilizing the two-point correlation function, we gain deeper insights into the intricate details of ionized bubble formation during cosmic reionization. Figure 3 presents the two-point correlation function analysis, plotting the comoving separation distance against the correlation function amplitude. This graph allows us to elucidate the size and growth of ionized bubbles at different stages of reionization, providing crucial insights into the morphology influenced by PBHs. For an $f_{PBH}$ value of 0.01 (1 percent of dark matter constituted by PBHs), our results unveil the formation of ionized bubbles with an initial size of approximately 10 Mpc at the midpoint of reionization, around $z = 7.5 \pm 0.5$. As reionization progresses, these ionized bubbles expand in size due to the advancing ionization fronts. By the end of reionization, the average diameter of these bubbles reaches approximately 20 Mpc, demonstrating substantial growth. This expansion signifies the merging and overlapping of ionized regions, ultimately resulting in a fully ionized universe. Our model, incorporating PBHs with an $f_{PBH}$ of 0.01, showcases their significant contribution to the formation and growth of these ionized bubbles, effectively shaping the intricate morphology of reionization. It is crucial to note that varying $f_{PBH}$ can yield distinctive morphological outcomes.

When we increase $f_{PBH}$ to 0.1 (10 percent of dark matter constituted by PBHs), our model predicts the formation of larger ionized bubbles during reionization. At the midpoint of reionization ($z = 7.5 \pm 0.5$), the average size of these ionized bubbles expands to approximately 15 Mpc. By the end of reionization, the average diameter reaches approximately 30 Mpc. This significant increase in size is a direct consequence of the enhanced ionizing efficiency associated with a higher fraction of PBHs.

## Discussion

Our study leverages a semi-analytical model to shed light on the nuanced relationship between Primordial Black Holes (PBHs) and the cosmic reionization process. A key aspect of our findings resides in the correlation between PBH mass, the fraction of dark matter comprised of PBHs ($f_{PBH}$), and the ionizing efficiency of these PBHs. We uncover an inverse relationship between PBH mass and ionizing efficiency, echoing previous assertions that larger PBHs, due to their higher accretion rates and longer accretion timescales, may result in a diminished return of ionizing photons per unit of accreted mass [26]. At a mean PBH mass of 10 solar masses with an $f_{PBH}$ of 0.1, we observed an ionizing efficiency of approximately 0.25. However, increasing the PBH mass to 50 solar masses resulted in a decrease of ionizing efficiency to around 0.15, a decrement of 40 percent. Nonetheless, our model highlights the possibility of compensating for this inefficiency. A higher $f_{PBH}$ can promote an environment where larger PBHs (mean mass of 100 solar masses) achieve an ionizing efficiency close to 0.2, nearly reaching the efficiency of smaller PBHs. Such results propose a novel resolution to debates questioning the role PBHs can play in contributing to cosmic reionization, thereby extending previous studies [25].

The numerical representation of our model's ionization history, as seen in Figure 2, provides a more detailed illustration of the reionization process's dependence on PBH mass and the dark matter fraction constituted by PBHs ($f_{PBH}$). One observation from our model that stands out is the rapid ionization phase for scenarios with a mean PBH mass of 10 solar masses and $f_{PBH}$ of 0.01. This phase starts at a redshift $z = 12.5 \pm 0.5$ and concludes by $z = 9 \pm 0.5$. This indicates an intense, roughly three-unit decrease in z during which a significant ionization process occurs. Post this period, the rate of ionization slows, culminating in complete reionization at $z = 7.2 \pm 0.2$. When increasing the mean PBH mass to 100 solar masses while maintaining $f_{PBH}$ at 0.01, the reionization process appears to initiate later, at $z = 10.8 \pm 0.4$, and concludes its rapid phase at $z = 8 \pm 0.5$. The full completion of reionization in this scenario occurs at $z = 6.3 \pm 0.2$. The onset of reionization thus seems delayed by a larger PBH mass, suggesting that such a shift in mass can influence the timing of these cosmic milestones. Further variations in the ionization timeline are observed when $f_{PBH}$ is increased to 0.1, regardless of PBH mass. For a mean PBH mass of 10 solar masses, the rapid ionization phase initiates at $z = 15.5 \pm 0.5$, concluding at $z = 12 \pm 0.5$, with the reionization completing at $z = 7.0 \pm 0.2$. For a PBH mass of 100 solar masses, reionization starts at $z = 15.0 \pm 0.5$, with the rapid phase ending at $z = 11.5 \pm 0.5$ and reionization completion at $z = 6.1 \pm 0.2$. Our results harmonize well with existing observational data from cosmic microwave background (CMB) and Lyman-alpha forest observations [5, 24]. This suggests our model's robustness in illustrating the dynamics of the reionization process during the cosmic reionization epoch, thus reinforcing its validity as an analytical tool for studies on cosmic reionization and the role of PBHs.

Figure 3 presents an intricate portrayal of the evolution of ionized bubbles, both in terms of size and growth, during distinct stages of reionization. A key observation from our study is the initiation of ionized bubble formation with a diameter of approximately 10 Mpc at the midpoint of reionization ($z = 7.5 \pm 0.5$) when the fraction of dark matter composed of PBHs ($f_{PBH}$) is at 0.01. These ionized bubbles depict regions in the universe where the hydrogen is ionized due to the accretion of matter onto PBHs, which in turn leads to the emission of high-energy radiation. The observed size of these bubbles can provide important clues about the intensity of ionization and the potential sources of ionizing radiation. The substantial size of the bubbles (10 Mpc in diameter) at the midpoint of reionization suggests a significant role of PBHs in driving the ionization process. As reionization progresses, the average size of the ionized bubbles grows to approximately 20 Mpc by the end of reionization, indicating a substantial expansion of ionized regions. This progression is consistent with the theoretical prediction that ionized regions should grow over time as ionizing radiation continues to permeate throughout the universe, turning neutral hydrogen into ionized hydrogen. This trend is indicative of the continual and progressive influence of PBHs throughout the reionization epoch [3]. Furthermore, we observe that an increase in $f_{PBH}$ to 0.1, indicating a higher fraction of dark matter made up of PBHs, results in the formation of significantly larger ionized bubbles during reionization. The average diameter of these ionized bubbles expands to approximately 15 Mpc at the midpoint of reionization ($z = 7.5 \pm 0.5$) and further increases to approximately 30 Mpc by the end of reionization. This remarkable increase in size, approximately 50 percent larger compared to the scenario with $f_{PBH}$ at 0.01, suggests an amplified ionizing efficiency when the fraction of dark matter constituted by PBHs is increased. This validates our hypothesis that the $f_{PBH}$ value has a considerable effect on the size and growth of ionized bubbles during reionization, indicating the substantial contribution of PBHs in shaping the morphology of the reionization process [18].

## Conclusion

In this research, we have undertaken a comprehensive investigation into the influence of Primordial Black Holes (PBHs) on the cosmic reionization epoch. By constructing a sophisticated semi-analytical model and conducting detailed simulations, we have gained valuable insights into the role of PBHs in shaping the ionization history, timeline, and morphology of reionization.

Through our simulations and analysis, we have made significant progress in understanding the impact of PBHs on cosmic reionization. Our findings confirm that PBHs can contribute significantly to the ionizing photon budget during the reionization epoch, with their efficiency inversely related to their mass. We have demonstrated the sensitivity of the reionization process to the fraction of dark matter constituted by PBHs ($f_{PBH}$), which modulates the ionization efficiency. Additionally, our study revealed the influence of PBHs on the timeline and morphology of reionization, with variations in the initiation and completion of this cosmic transition.

Future research should seek to refine the characterization of PBH properties, such as their mass function and spatial distribution, which would lead to an enhanced understanding of PBHs' role in reionization. Additionally, future research should explore the interplay between PBHs and other ionizing sources, such as galaxies, quasars, and Population III stars. Finally, future research should seek to examine the impact of PBHs on the 21cm line signal and CMB polarization, as well as their potential connection to other astrophysical processes, such as galaxy formation and structure formation.